\documentclass[sigconf]{acmart}

\usepackage{lineno,hyperref,listings,subfig,tabularx,color, tikz, xcolor, tcolorbox, graphicx}
\usepackage[skip=4pt]{caption}

\graphicspath{{figures/}}

\setcopyright{none}
\renewcommand\footnotetextcopyrightpermission[1]{}

\DeclareFixedFont{\ttb}{T1}{txtt}{bx}{n}{8} 
\DeclareFixedFont{\ttm}{T1}{txtt}{m}{n}{8}  
\usepackage{color}
\definecolor{deepblue}{rgb}{0,0,0.5}
\definecolor{deepred}{rgb}{0.6,0,0}
\definecolor{deepgreen}{rgb}{0,0.5,0}

\newcommand\pythonstyle{\lstset{
language=Python,
basicstyle=\ttm,
otherkeywords={self},             
keywordstyle=\ttb\color{deepblue},
emph={MyClass,__init__},          
emphstyle=\ttb\color{deepred},    
stringstyle=\color{deepgreen},
frame=tb,                         
showstringspaces=false            %
}}

\lstnewenvironment{python}[1][]
{
\pythonstyle
\lstset{#1}
}
{}

\title{Triggerflow: Trigger-based Orchestration of Serverless Workflows}
\AtBeginDocument{%
  \providecommand\BibTeX{{%
    \normalfont B\kern-0.5em{\scshape i\kern-0.25em b}\kern-0.8em\TeX}}}

\begin{document}

\author{Pedro~Garc\'ia L\'opez}
\affiliation{%
  \institution{IBM Watson Research}
  \city{New York}
  \country{USA}
}
\email{pedro.garcia.lopez@ibm.com}

\author{Aitor Arjona}
\affiliation{%
  \institution{Universitat Rovira i Virgili}
    \city{Tarragona}
  \country{Spain}
}
\email{aitor.arjona@urv.cat}

\author{Josep~Samp\'e}
\affiliation{%
  \institution{Universitat Rovira i Virgili}
    \city{Tarragona}
  \country{Spain}
}
\email{josep.sampe@urv.cat}

\author{Aleksander Slominski}
\affiliation{%
  \institution{IBM Watson Research}
    \city{New York}
  \country{USA}
}
\email{aslom@ibm.com}

\author{Lionel Villard}
\affiliation{%
  \institution{IBM Watson Research}
    \city{New York}
  \country{USA}
}
\email{villard@us.ibm.com}

\begin{abstract}

As more applications are being moved to the Cloud thanks to serverless computing, it is increasingly necessary to support native life cycle execution of those applications in the data center. 

But existing systems either focus on short-running workflows (like IBM Composer or Amazon Express Workflows) or impose considerable overheads for synchronizing massively parallel jobs (Azure Durable Functions, Amazon Step Functions, Google Cloud Composer). None of them are open systems enabling extensible interception and optimization of custom workflows.

We present Triggerflow: an extensible  Trigger-based Orchestration architecture for serverless workflows built on top of Knative Eventing and Kubernetes technologies. We demonstrate that Triggerflow is a novel serverless building block capable of constructing different reactive schedulers (State Machines, Directed Acyclic Graphs, Workflow as code). We also validate that it can support high-volume event processing workloads, auto-scale on demand and transparently optimize scientific workflows.

\end{abstract}

\keywords{event-based, orchestration, serverless}

\fancyhead{
     \vspace{-30pt}
     \begin{tikzpicture}
         \node[align=center] () at (0,0) {
             \begin{tcolorbox}[colback=white!100,colframe=black!50,width=\textwidth,sharp corners]
						 \small\centering CC-BY 4.0. This is the author's preprint version of the camera-ready article. A full version of this paper is published in the proceedings of 14th ACM International Conference on Distributed and Event-Based Systems (DEBS 2020).
             \end{tcolorbox}
         };
     \end{tikzpicture}
}

\maketitle

\thispagestyle{fancy}

\section{Introduction}

Serverless Function as a Service (FaaS) is becoming a very popular programming model in the cloud thanks to its simplicity, billing model and inherent elasticity. The FaaS programming model is considered event-based, since functions are activated (triggered) in response to specific Cloud Events (like a state change in a disaggregated object store like Amazon S3).

The FaaS model has also proven ideally suited (PyWren \cite{pywren}, ExCamera \cite{fouladi}) for executing embarrassingly parallel computing tasks. But both PyWren and ExCamera required their own ad-hoc external orchestration services to synchronize the parallel executions of functions. For example, when the PyWren client launches a map job with N functions, it waits and polls Amazon S3 until all the results are received in the S3 bucket. ExCamera also relied on an external Rendezvous server to synchronize the parallel executions.

Lambda creator Tim Wagner recently outlined \cite{wagner} that Cloud providers must offer new serverless building blocks to applications. In particular, he foresees new services like fine-grained, low-latency orchestration, execution data flows,  and the ability to customize code and data at scale to support the emerging data-intensive applications over Serverless Functions.

The reality is that existing serverless orchestration systems are not designed for long-running data analytics tasks \cite {wosc4,wosc5}. Either they are focused on short-running highly interactive workflows (Amazon Express Workflows, IBM Composer) or impose considerable overheads for synchronizing massively parallel jobs (Azure Durable Functions, Amazon Step Functions, Google Cloud Composer).

We present Triggerflow, a novel building block for composing event-based services. As more applications are moved to the Cloud, this service will enable to control the life-cycle of those applications in a reactive and extensible way. The flexibility of the system can also be used to transparently optimize the execution of tasks in reaction to events.

The major contributions of this paper are the following:
\begin{enumerate}

\item We present a Rich Trigger framework following an Event-Condition-Action (ECA) architecture that is extensible at all levels (Event Sources and Programmable Conditions and Actions). Our architecture ensures that composite event detection and event routing mechanisms are mediated by reactive event-based middleware.

\item We demonstrate Triggerflow's extensibility and universality creating atop it a state machine workflow scheduler, a  DAG engine, an imperative Workflow as Code (using event sourcing) scheduler, and integration with an external scheduler like PyWren. We also validate performance and overhead of our scheduling solutions compared to existing Cloud Serverless Orchestration systems like Amazon Step Functions, Amazon Express Workflows, Azure Durable Functions and IBM Composer.

\item We finally propose a generic implementation of our model over standard CNCF Cloud technologies like Kubernetes, Knative Eventing and CloudEvents. We validate that our system can support high-volume event processing workloads, auto-scale on demand and transparently optimize scientific workflows. The project is available as open-source in \cite{triggerflow}.

\end{enumerate}


\newpage
\section{Related work}

FaaS is based on the event-driven programming model.  In fact, many event-driven abstractions like triggers,  Event Condition Action (ECA) and even composite event detection were already inspired by the veteran Active Database Systems \cite{active}. 


Event-based triggering has also been extensively employed in the past to provide  reactive coordination of distributed systems \cite{mitchell2012oolong, han2013large}.  Event-based mechanisms and triggers have also been extensively used \cite{eve, chen2008,binder2006, padres} in the past to build workflows and orchestration systems. The ECA model including trigger and rules fits nicely to define the transitions of finite state machines representing workflows. In  \cite{dai2018trigger}, they propose to use synchronous aggregation triggers to coordinate massively parallel data processing jobs.

An interesting related work is \cite{padres}.  They leverage composite subscriptions in content-based publish/subscribe systems to provide decentralized Event-based Workflow Management.  Their PADRES system supports parallelization, alternation, sequence, and repetition compositions thanks to content-based subscriptions in a Composite Subscription Language. 

More recently, a relevant article \cite{cep} has surveyed the intersections of the Complex Event Processing (CEP) and Business Process Management (BPM) communities. They clearly present the existing challenges to combine both models and describe recent efforts in this area. We outline that our paper is in line with their challenge ``Executing business processes via CEP rules", and our novelty here is our serverless reactive and extensible architecture.

In serverless settings,  the more relevant related work aiming to provide reactive orchestration of serverless functions is the Serverless trilemma \cite{trilemma} from IBM. In their paper, the authors advocate for reactive run-time support for function orchestration, and present a solution for sequential compositions on top of Apache OpenWhisk. 


A plethora of academic works are proposing different so-called serverless orchestration systems like  \cite{wukong, ripple,  malawski, formal, specrg,gg}. However, most of them rely on non-serverless services like VMs or dedicated resources, or they use functions calling functions patterns which complicate their architectures and fault tolerance.  None of them offer extensible trigger abstractions to build different schedulers.

All Cloud providers are now offering cloud orchestration and function composition services like IBM Composer, Amazon Step Functions, Azure Durable Functions, or Google Cloud Composer.

IBM Composer service is in principle designed for short-running synchronous composition of serverless functions. IBM Composer generates a state machine representation of the workflow  to be executed with IBM Cloud Functions. It can represent sequences, conditional branching, loops, parallel, and map tasks. However,  fork/join synchronization (map, parallel)   blocks on an external user-provided Redis service, limiting their applicabillity to short running tasks.



Amazon offers two main services: Amazon Step Functions (ASF) and Amazon Step Functions Express Workflows (ASFE). The Amazon States Language (based on JSON) permits to model task transitions, choices, waits, parallel, and maps in a standard way. ASF is a fault-tolerant managed service designed to support long-running workflows and ASFE is designed for short-running (less than five minutes) highly intensive workloads with relaxed fault-tolerance.

Microsoft's Azure Durable Functions (ADF) represents workflows as code using C\# or Javascript,  leveraging async/await constructs and using event sourcing to replay workflows that have been suspended.  ADF does not support map jobs explicitly, and only includes a \emph{Task.whenAll} abstraction enabling fork/join patterns for a group of asynchronous tasks.
 
Google offers Google Cloud Composer service leveraging a managed Apache Airflow cluster.  Airflow represents workflows in a DAG (Directed Acyclic Graph) coded in Python, so that it cannot support cycles. It is not ideally suited for parallel jobs or high-volume workflows, and it is not designed for orchestrating serverless functions.

Two previous papers \cite{wosc4,wosc5} have compared public FaaS orchestration services for coordinating massively parallel workloads.  In those studies, IBM Composer offered the fastest performance and reduced overheads to execute map jobs whereas ASF or ADF imposed considerable overheads.  We will also show in this paper how ASFE obtains good performance for parallel workloads.



None of the existing cloud orchestration services is offering an open and extensible trigger-based API enabling the creation of custom workflow engines.  We demonstrate in this paper that we can use Triggerflow to implement existing models like ASF or Airflow DAGs. Triggerflow is not just another scheduler, but a reactive meta-tool to build reactive schedulers leveraging Knative standard technologies.


\subsection{Cloud Event Routing and Knative Eventing}

Event-based architectures are gaining relevance in Cloud providers as a unifying infrastructure for heterogeneous cloud services and applications.  Event services participate in the entire cloud control loop from event production in event sources, to event detection using monitoring services, to event logging and data analytics of existing event workflows, and finally to service orchestration and event reaction thanks to appropriate filtering mechanisms.

The trend is to create cloud event routers, specialized rule-based multi-tenant services, capable of filtering and triggering selected targets in the Cloud in response to events.  Amazon is offering EventBridge, Azure offers EventGrid, and Google and IBM are investing in the open Knative Eventing project and CNCF CloudEvents standard. 



The Knative project  was created to provide streamlined serverless-like experience for developers using Kubernetes. It contains a set of high-level abstractions related to scalable functions (Knative Serving) and event processing (Knative Eventing) that allows the description of asynchronous, decoupled, event-driven applications built out of event sources, sinks, channels, brokers, triggers, filters, sequences, etc.


The goal of Knative is to allow developers to build cloud native event-driven serverless applications on those abstractions. The value of Knative is to encapsulate well tested best practices in high-level abstractions that are native to Kubernetes: custom resource definitions (CRDs) for new custom resources (CRs) such as event sources. Abstractions allow developers to describe event-driven application components and have late-binding to underlying (possibly multiple) messaging and eventing systems like Apache Kafka and NATS among others.



Triggerflow aims to leverage existing event routing technology (Knative Eventing) to enable extensible trigger-based orchestration of serverless workflows. Triggerflow includes advanced abstractions not present in Knative Eventing like dynamic triggers, trigger interception, custom filters, termination events, and a shared context among others. Some of these novel services may be adopted in the future by event routing services to make it easier to compose, stream,  and orchestrate tasks.

\section{Triggerflow Architecture}

\medskip
\begin{figure*}[ht!]
	\includegraphics[width=0.9\textwidth]{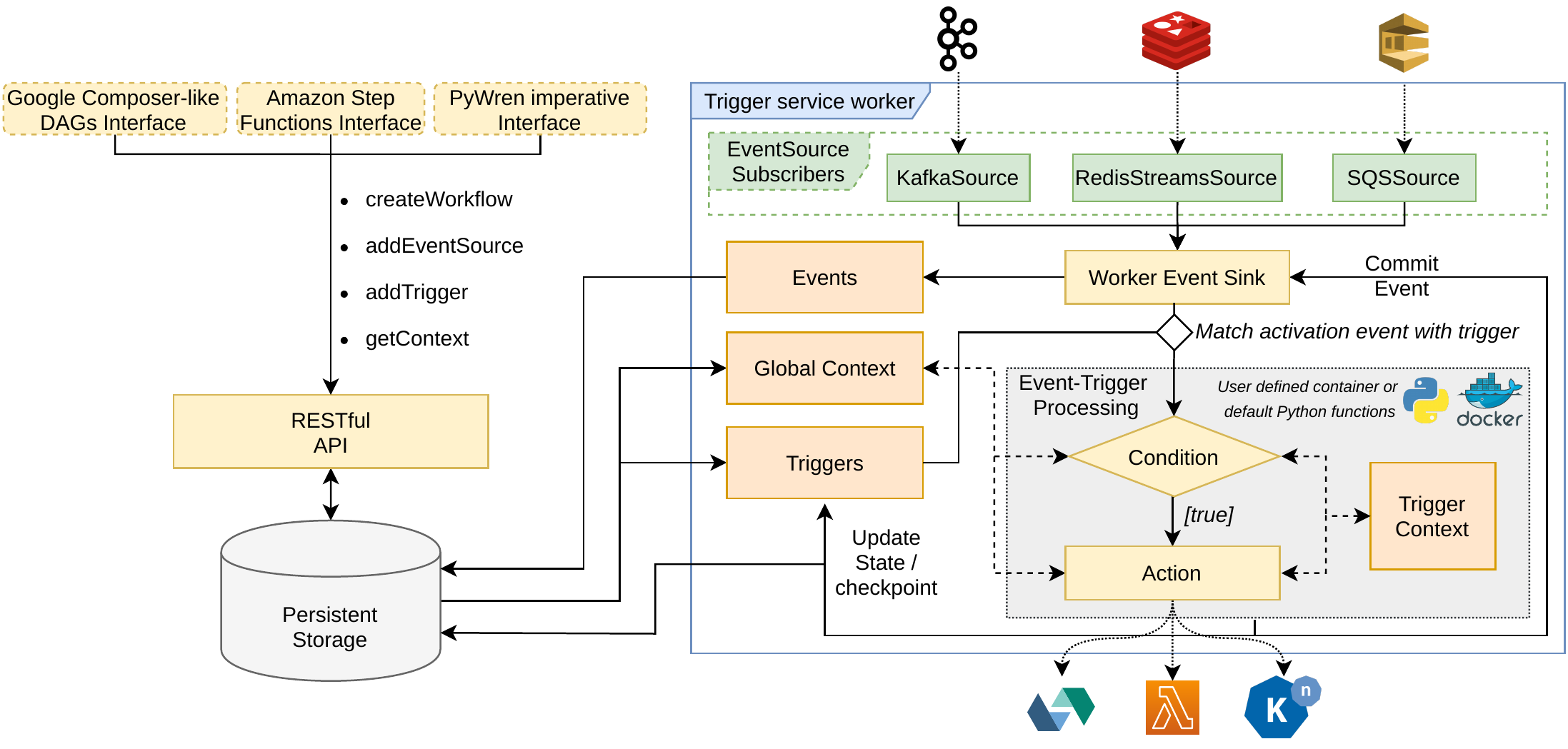}
	\centering
	\caption{Triggerflow Architecture}
	\label{fig:Triggerflow_architecture}
	\vspace{-8pt}
\end{figure*}

We can see in Figure~\ref{fig:Triggerflow_architecture} an overall diagram of the Triggerflow Architecture. The Trigger service follows an extensible Event-Condition-Action architecture.  The service can receive events from different Event Sources in the Cloud (Kafka, RabbitMQ, Object Storage, timers).  It can execute different types of Actions  (containers, Functions, VMs). And it can also enable the creation of custom filters or Conditions from third-parties. The Trigger service also provides a shared persistent context repository providing durability and fault tolerance.

Figure~\ref{fig:Triggerflow_architecture} also shows the basic API exposed by TriggerFlow: \textit{createWorkflow} initializes the context for a given workflow, \textit{addTrigger}  adds a  new trigger (including event, conditions, actions, and context), \textit{addEventSource} permits the creation of new event sources, and \textit{getState} obtains the current state associated to a given trigger or workflow. 

Different applications and schedulers can benefit from serverless awakening and rich triggering by using this API to build different orchestration services like Airflow-like DAGs, ASF state machines or Workflow as Code clients like PyWren.

\subsection{Design goals}

Let's establish a number of design goals that must be supported in the proposed architecture:

\begin{enumerate}

\item Support for Heterogeneous Workflows:  The main idea is to build a generic building block for different types of schedulers. The system should support enterprise workflows based on Finite State Machines, Directed Acyclic Graphs, and Workflow as Code systems.  

\item Extensibility and Computational Reflection:  The system must be extensible enough to support the creation of novel workflow systems with special requirements like specialized scientific workflows. The system must support introspection and interception mechanisms enabling the monitoring and optimization of existing workflows.

\item Serverless design:  The system must be reactive, and only execute logic in response to events, like state transitions. Serverless design also entails pay per use, flexible scaling, and dependability. 



\item Performance: The system should support high-volume workloads like data analytics pipelines with numerous parallel tasks. The system should exhibit low overheads for both short-running and long-running workflows.

\end{enumerate}

\subsection{Trigger service}

Our proposal is to design a purely event-driven and reactive  architecture for workflow orchestration.  Like previous works \cite{eve, chen2008,binder2006}, we also propose to handle state transitions using  event-based triggering mechanisms. The novelty of our approach precisely relies on the aforementioned design goals: support for heterogeneous workflows, extensibility, serverless design, and performance for high volume workloads.

We follow an \textbf{Event Condition Action} architecture in which triggers (active rules) define which action must be launched in response to Events or to Conditions evaluated over one or more Events. The system must be extensible at all levels: Events, Conditions, and Actions. Let us introduce some definitions for our event-based orchestration model. 

\textbf{Definition 1.  Workflow:} We can represent a workflow as a Finite State Machine (FSM)  being a 6-tuple with 
\newline
 M = ($\sum_{in},Ctx, S, s,F, \delta$), 
in this 6-tuple:
\begin{enumerate} 
\item  $\sum_{in}$: the set of input events
\item  Ctx: the set of context variables
\item  S: the set of states which map to Actions in the ECA model
\item  s: initial state, linked to an initial event
\item  F: end state, linked to a final Termination event
\item  $\delta$: state-transition function: $\delta:  S \times \sum  \rightarrow S $ , based on the ECA triggers
\end{enumerate}

\textbf{Definition 2. Trigger ($\delta$):} can be defined as the state transition function. The trigger is a 4-tuple with 
(Event, Context, Condition, Action) that  moves one state to the following when the condition on input events holds. In this case, the trigger launches the appropriate action which corresponds to the next state.  Each action will in turn fire events that may be captured by another trigger. Triggers can be transient and dynamic (activated on demand) or persistent if they remain always active.  Its components are:

\begin{itemize}
  \item \textbf{Event}: Events are the atomic piece of information that drive flows in Cloud applications. We rely on the standard CNCF CloudEvents version 1.0 specification  to represent events.  To match an event to its trigger, the \textit{subject} and \textit{type} fields of a CloudEvent are used.  We use the  \textit{subject} field to match the event to its corresponding trigger, and the \textit{type} field to describe the type of the event. Termination and failure events use this \textit{type} field  to notify success or failure.

  \item \textbf{Context}: The context is a fault-tolerant key-value data structure that contains the state of the trigger during its lifetime. It is also used to introspect the current trigger deployment, to modify the state of other triggers or to dynamically activate/deactivate triggers. 
  \item \textbf{Condition}: Conditions are active rules (user-defined code) that filter events to decide if they match in order to launch the corresponding action.  Conditions evaluate rules over  primitive  events (single)  or over composite (group) events. Composite event information like counters may be stored in the Context.  Conditions produce a \textit{boolean} result that represents whether the trigger has to be fired or not.
  
  \item \textbf{Action}: Actions are the computations (user-defined code) launched in response to matching Conditions in a trigger.  An Action can be a serverless function, VM or container. When it is executed, we consider that the trigger has been fired.  
\end{itemize}

\textbf{Definition 3. Mapping workflow to triggers:} A workflow can be mapped to a set of Triggers ($\Delta$) which contains all state transitions ($\delta$ triggers) in the State Machine.  

We will show in next sections how different workflows (Amazon Step Functions) and Directed Acyclic Graphs (Apache Airflow)  can be transformed to a set of triggers ($\Delta$), which is the information needed by the Trigger service to orchestrate them.
For example,  to transform a DAG into triggers, a trigger is added for every edge (workflow transition) of the graph. In a DAG, every node has its own unique ID, so the termination event from a task will contain as subject its ID to fire the trigger that handles its termination and invokes the next step in the workflow.

\textbf{Definition 4. Substitution principle:} A Workflow  must comply with an Action according to triggering (initialization) and finalization (Termination Event). A homogeneous treatment of Workflows and Actions permits nested workflow composition and iterations.








\textbf{Definition 5.  Dynamic Trigger interception:} Any trigger can be intercepted dynamically and transparently to execute a desired action. Interception code is also performed with triggers. It must be possible to intercept triggers by condition identifier or by trigger identifier.  The condition identifier represents each existing condition in Triggerflow, for example a  map condition that aggregates all events in a parallel invocation. The trigger identifier represents the unique ID that each trigger receives on creation. 


We can introspect workflows, triggers, conditions, and actions using the Context. And we can intercept any trigger in the system in a transparent way using the Rich Trigger API. This opens the system to customize code and data in a very granular way. 

\section{Prototype Implementation}
\label{sec:implementation}

We have developed two different implementations of Triggerflow: one over Knative, which follows a push-based mechanism to pass the events from the event source to the appropriate worker, and another one using Kubernetes Event-driven Autoscaling (KEDA), where the worker follows a pull-based mechanism to retrieve the events directly from the event source. We created the prototypes on top of the IBM Cloud infrastructure, leveraging the services in its catalog to deploy the different components of our architecture. These components are the following:

\begin{itemize}
\item A Front-end RESTful API, where a user connects to interact with Triggerflow.
\item A Database, responsible for storing workflow information, such as triggers, context, etc.
\item A Controller, responsible for creating the workflow workers in Kubernetes.
\item The workflow workers (TF-Worker hereafter), responsible for processing the events by checking the triggers' conditions, and applying the actions.
\end{itemize} 

In our implementation, each workflow has its own TF-Worker. In other words, the scalability of the system is provided at workflow-level and not at TF-Worker level. In the validation (Sec. \ref{sec:validation}), we demonstrate how each TF-Worker  provides enough event ingestion rate to process large amounts of events per second. 



In our system, the events are logically grouped in what we call \emph{workflows}. The \emph{workflow} abstraction is useful, for example, to differentiate and isolate the events from multiple workflows, allowing to share a common context among the (related) events.

\subsection{Deployment on Knative}
\label{sec:deployment_Knative}

We mainly benefit from the Knative auto scaler component in Knative Serving and the routing/filtering service in Knative Eventing.

Any serverless reactive architecture requires a managed multi-tenant component that is constantly running, monitoring event sources, and only launching actions in response to specific events. In this way, the tenant only pays for the execution of actions in response to events, and not for the constant monitoring of event services. For example, in OpenWhisk, when we create a trigger for a Function (like an Object Storage trigger), the system is in charge of monitoring the event source and only launching the function in response to events. 

In Knative Eventing, each tenant will have an Event Source that receives all events they are interested in (and have access to). We register a Knative Eventing trigger for each workflow in the system. The filtering capabilities of Knative Eventing's trigger permit to route events of this workflow to the appropriate TF-Worker (Condition).  

Each workflow event is tagged with a unique workflow identifier. We have created a customized functions runtime, which generates function termination events to the desired message broker that include the selected workflow identifier. If Triggerflow must receive events from services which do not include this workflow ID, a generic filtering service will match conditions to the incoming event (like ``all events of this object storage bucket belong to this workflow"), tag the event, and route it to the tenant's Event Source.

As each event contains a unique identifier per workflow, it is easy for Knative eventing to route this event to the selected TF-Worker.  The TF-Worker is then launched by Knative Serving to process the event, but it will also scale to zero if no more events are produced in a period. This ensures the serverless scale to zero and pay as you go qualities for our Triggerflow service. The TF-Worker accesses workflow state in the Context persistent store, which is also used for checkpointing and fault tolerance. 

Regarding \textbf{fault tolerance}, Knative Eventing guarantees "at least once" message delivery, and automatic detection and restart of failed workers.  If a TF-Worker fails, the persistent Context will restore the state in a consistent manner after the failure. The persistent Context is also used for stateful Conditions, like aggregation fork-join triggers that perform composite event detection and event counting.


\begin{figure}[t!]
	\includegraphics[width=0.48\textwidth]{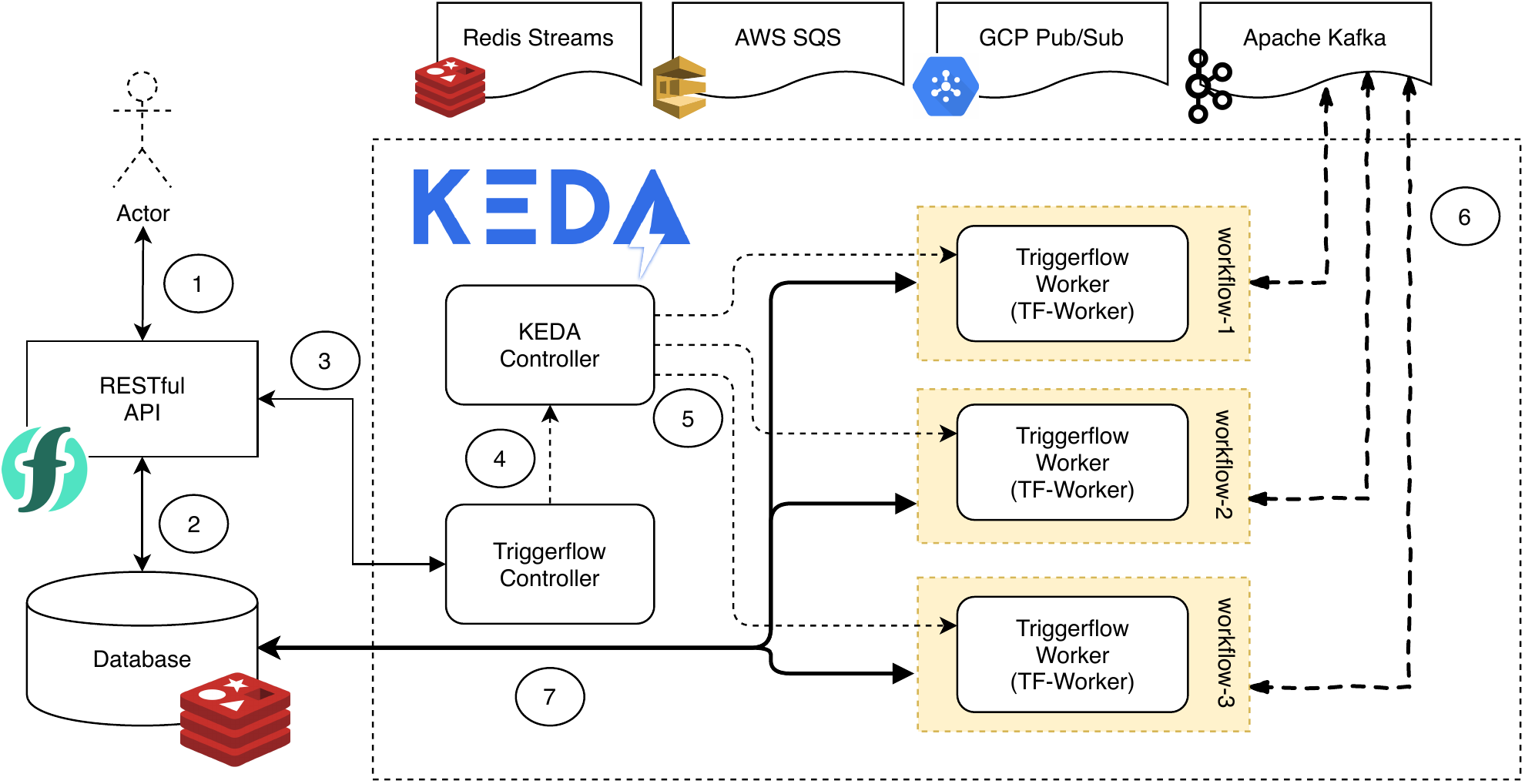}
	\centering
	\caption{Prototype deployment on KEDA}
	\label{fig:impl_keda}
	\vspace{-8pt}
\end{figure}

\subsection{Deployment on KEDA}

One of the hardest problems in event-driven applications is to deal with reliability and scalability. Event systems may be receiving events as soon as they are created ("pushed") or they may process them when they are ready ("pull" or "poll") and for both cases they need to deal with capacity limits and error handling. Knative is very well suited for push-based scaling as it can auto-scale based on incoming HTTP requests containing events.  Kubernetes Event-driven Autoscaling (KEDA) is the best option now for event-based configurable pull-based scaling.

We have also  implemented Triggerflow entirely on top of  Kubernetes using the KEDA project  \cite{keda}. KEDA offers pull-based configurable event queue monitoring and reactive scalable instantiation of Kubernetes containers. KEDA also offers configurable auto-scaling mechanisms to scale up or down to zero.

In this case, the  Triggerflow Controller integrates KEDA for the monitoring of Event Sources and for launching the appropriate TF-Workers, and scaling them to zero when necessary. It is also possible to configure different parameters in KEDA like the queue pulling interval, passivation interval, and number of events scaling interval. Different types of workflows may require different configuration parameters.

The advantage here is that, unlike in Knative Eventing, our TF-Workers connect directly to the Message Broker (Kafka, Redis Streams) using the native protocol of the broker. This permits to handle more events per second in a single pod. As we demonstrate in the validation, this allows us to handle intensive workloads from scientific workflows coordinating parallel jobs over thousands of serverless functions.



Figure \ref{fig:impl_keda} shows a high-level perspective of our implementation using KEDA. In this deployment, Triggerflow works as follows:  Through the client, a user must firstly create an empty workflow to the Triggerflow registry, and reference an event source that this workflow will use. Then, the user can start adding triggers to it (1). All the information is persisted in the database (for example, Redis) (2). Then, immediately after creating the \emph{workflow}, the front-end API communicates with the \emph{Triggerflow controller} (3), deployed as a single stateless pod container (service) in Kubernetes, to create the auto-scalable \emph{TF-Worker} in KEDA (4). From this moment, KEDA is responsible to scale up and down the TF-Workers (5). In KEDA, as stated above, the TF-Worker is responsible for communicating directly to the event source (6) to pull the incoming events. Finally, TF-Workers periodically interact with the database (7) to keep the local cache of available triggers updated, and to store the context (checkpointing) for fault-tolerance purposes.

Regarding \textbf{fault tolerance}, we also guarantee "at least once" message delivery and restarting of failed workers. In this case,  the TF-Worker uses batching to commit groups of events in the Kafka Event Source once they have been correctly processed. If the TF-Worker fails, Kafka will just resend the non-committed events to the TF-Worker and thus ensuring message delivery.

In our Redis implementation, we use Redis both as event broker (Redis Streams), and  as persistent store (for the Context and events). Again, if the TF-Worker fails, all events are in the event store, so it will continue with the non-processed events.

If Knative Eventing and KEDA communities converge in the next months, we will be able to deploy Triggerflow directly on top of one unified event router technology.  It is also possible that some building blocks of Triggerflow could be moved to the Knative Eventing kernel. For example, the Knative Eventing community is now considering more advanced filtering mechanisms (complex event processing). In that case, our TF-Worker could delegate many tasks to the underlying event router.

\section{Use cases}
\label{sec:use_cases}

To demonstrate the flexibility that can be achieved using triggers with programmable conditions and actions, we have implemented three different workflow models that use Triggerflow as the underlying serverless and scalable workflow orchestrator.

\subsection{Directed Acyclic Graphs}

When a workflow is described as a Directed Acyclic Graph (DAG), the vertices of the graph represent the tasks of the workflow and the edges represent the dependencies between the tasks. The fact that a DAG does not have cycles implies that there are no cyclic dependencies, which would be impossible to fulfill. 

The orchestration platforms that rely on DAGs for their workflow description, such as Apache Airflow, handle the dependencies between tasks with their \textit{downstream relatives} attribute, i.e. upon a completion of a task execution, these orchestrators look for what tasks have to be executed after the completed task.

However, from a trigger-based orchestration perspective, it is more compelling to know what tasks have to be executed before a certain one, i.e. what are the dependencies of every task, their \textit{upstream relatives}. With this information, we can register a trigger to activate a task's execution when all termination events from its upstream relatives are present.

To orchestrate a workflow defined as a DAG with triggers, we will define a trigger for every edge of the DAG:
\begin{itemize}
	\item As \textbf{activation events} of the trigger, we register the task IDs that have to be completed before the tasks that the edge points to (their \textit{upstream relatives}).
	\item As \textbf{condition}, we count the number of events the trigger has to aggregate before executing the next task (i.e. a join of a map execution).
	\item As \textbf{action}, we register the actual task to be executed, ideally an asynchronous task such as an invocation of a serverless function.
\end{itemize}

\begin{figure}[t!]
	\includegraphics[width=0.4\textwidth]{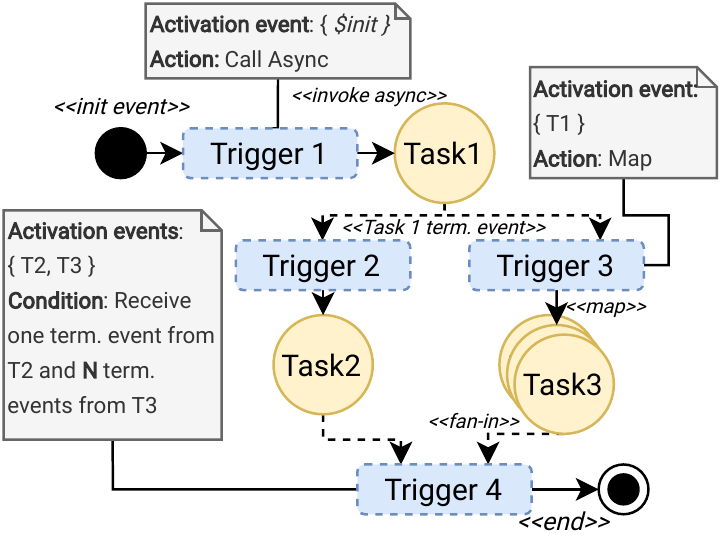}
	\centering
	\caption{Triggers that connect the tasks of an example DAG}
	\label{fig:dag2triggers}
	\vspace{-8pt}
\end{figure}

To orchestrate a workflow in this way, it is assumed that after an asynchronous task is completed, it will produce a termination event containing its ID to activate the trigger that manages the task execution that follows it.

To handle a map-join trigger condition, before actually making the invocation requests, we use the \textit{introspect} context feature from the activated trigger action to dynamically modify the condition of the trigger that will aggregate the events, to set the specific number of expected functions to be joined. This is used in the case that the iterator which we map onto has a variable length depending on the workflow execution.

Furthermore, this approach gives us the opportunity to handle errors during a workflow runtime. Special triggers can be added that activate when a task fails, so that the trigger action can handle the task's error and halt the workflow execution until the error is solved. After error resolution (retry, skip or try-catch logic), the workflow's execution can be resumed by activating the corresponding trigger that would have been executed in the first place, as if there had not been an error.


The DAGs interface implementation is inspired by Airflow's extensible DAG definition based on the  \textit{Operator} abstraction. According to Airflow's core ideas, an \textit{Operator} describes what is the actual work logic that is carried out by a task. Airflow offers a wide variety of operators to work with out of the box, but it can be extended through the implementation of \textit{plugins}. This approach is well suited to Triggerflow's architecture, thanks to its flexible programmatic trigger actions and conditions.

To illustrate this approach, Figure \ref{fig:dag2triggers} depicts how a simple DAG with \textit{call async}, \textit{maps}, and \textit{branches} is orchestrated using triggers.

\subsection{State Machines and Nested Workflows}

Amazon Step Functions bases its workflow description on a state machine defined by a declarative JSON object using the Amazon States Language DSL.

Similarly to Airflow's DAGs, a state machine definition in Amazon States Language (ASL) only takes into consideration what is the \textit{next} state to execute for each of them. However, from a trigger perspective, it is needed to figure out what states need to be executed before a given one, so that we can add a trigger that fires upon a state completion and executes the next one. Therefore, there will be a trigger for every state transition that handles the state machine flow logic.

Nevertheless, a distinctive feature that ASL provides is that a state can be a sub-state machine. For instance, the primitives \textit{map} and \textit{parallel}, map and branch to an entire state machine, rather than a single task like in the DAG interface. To manage this feature, we need a special event that is produced when a state machine ends. For map and branch joins, we will then join those sub-state machines instead of single tasks. To do so, we identify each sub-state machine with a unique tag in the scope of the execution. By doing so, we also comply with the substitution principle of the serverless trilemma.

To produce state machine termination events, we need to activate triggers from within a trigger action/condition function, as state machine joining is detected in there. To do so, the worker's event sink internal buffer was made accessible through the context object so that a trigger action/condition function can produce the events that activate the necessary subsequent triggers.

In an Amazon Step Functions execution, the states can transfer their output to the input of the following state. To reproduce this functionality, we use the \textit{Context} of the \textit{Workflow}, so that the output of a state can be saved in the trigger's context and accessed by other triggers.

If we consider a state machine to be itself a state, we can seamlessly compose ASL definitions in other state machines with its triggers and connections. Amazon Step Functions, however, is more limited in terms of task extensibility since we are given a closed set of state types. We will explain here how these are processed with triggers:

\begin{itemize}
    \item \textbf{Task} and \textbf{Pass} states: These state types carry out the actual workflow computational logic, the rest of the state types only manage the state machine flux. The Task state relies on the asynchronous Lambda invoked to signal the next trigger upon its termination, whereas the Pass state signals itself its termination event.
    \item \textbf{Choice} state: The choice state type defines a set of possible outcomes that execute depending on some basic boolean logic that can compare numbers, timestamps, and strings. The trigger approach for this state is simple: for all possible outcomes apply the condition defined in the Choice state to the condition field of the trigger that handles its state execution.
    \item \textbf{Parallel} state: This state type defines a set of sub-state machines that run in parallel. In this case, we will iterate each sub-state machine and collect their IDs. Finally, we add a trigger that is activated whenever any of those sub-state machines ends, but it is only executed when it has been signaled by every sub-state machine.
    \item \textbf{Map} state: Similarly to the Parallel state type, this state defines a single sub-state machine that executes for every element in an iterable data structure input in parallel. Before executing the sub-state machines, we first add a trigger that, during its action execution, checks the length of the iterable object (which is the number of parallel state machines, unknown until execution), and registers it to the trigger context that handles the sub-state machines termination stating how many of them it should wait for.
    \item \textbf{Wait} state: The Wait state type waits for a certain amount of seconds, or until a timestamp is reached before continuing. It can be implemented by registering the activation event production that activates the trigger to an external time-based scheduler.
    \item \textbf{Fail} and \textbf{Succeed} states: The Fail and Succeed states stop the execution of the state machine and determine if it executed successfully or failed. It can be implemented assigning special actions to their triggers that end the execution of the workflow.
\end{itemize}

\begin{figure}[ht]
	\includegraphics[width=0.4125\textwidth]{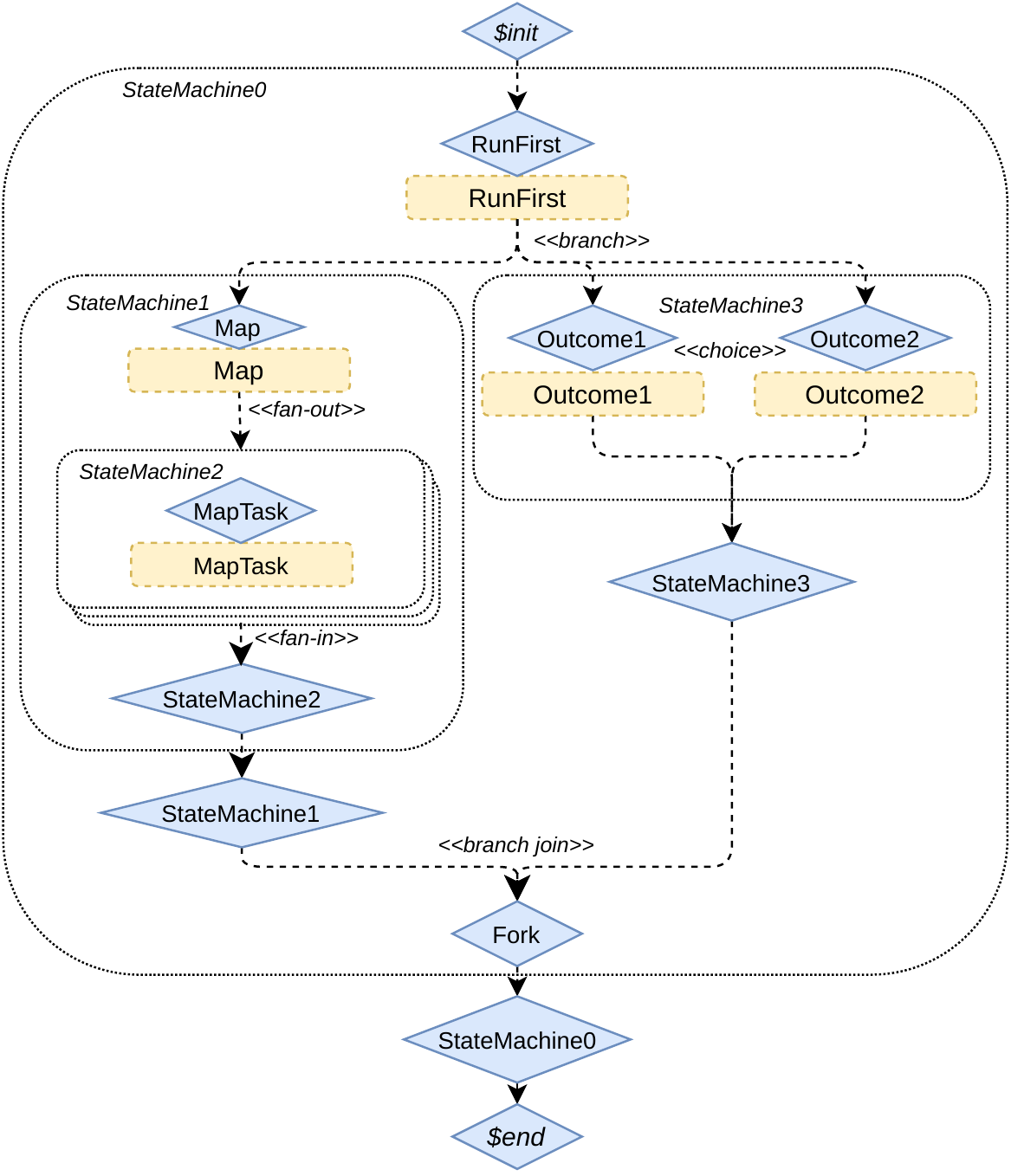}
	\centering
	\caption{Triggers representation of an ASF state machine}
	\label{fig:asf}
	\vspace{-8pt}
\end{figure}

Figure \ref{fig:asf} depicts how an ASF state machine is orchestrated by triggers.

\subsection{Workflow as Code and Event Sourcing}

The trigger service is also useful to reactively invoke an external scheduler because of  state changes caused by some condition. For example, Workflow as Code systems like PyWren or Azure Durable Functions represent state transitions as asynchronous function calls (async/await) inside code written in Python or C\#.  Asynchronous invocations and futures in PyWren or  async/await calls in Azure Durable Functions simplify code so developers can write synchronous-like code that suspends and continues when events arrive. 

The model supported by Azure Durable Functions is reactive and event-based, and it relies on event sourcing to restart the function to its current state.  We can use dynamic triggers to support external schedulers like Durable Functions that suspend their execution until the next event arrives.  For example, let's look at this PyWren code:

\begin{python}
import pywren_ibm_cloud as pywren

def my_function(x):
    return x + 7

pw = pywren.ibm_cf_executor()
future = pw.call_async(my_function, 3)
res = future.result()
futures = pw.map(my_function, range(res))
print(pywren.get_result(futures))
\end{python}

In this code, the functions \emph{call\_async} and \emph{map} are used to invoke one or many functions.
PyWren code like this is executed normally in the client in a notebook, which is usually adequate for short running workflows. But what if we want to execute a long-running workflow with PyWren in a reactive way? The solution is to run this PyWren code in Triggerflow reacting to events.  Here,  prior to perform any invocation, PyWren can register the appropriate triggers, for example:

\medskip
\textbf{call\_async(my\_function, 3)}: Inside this code we will dynamically register a function termination trigger.

\textbf{map(my\_function, range(res))}: Inside this code we will dynamically register an aggregate trigger for all functions in the map.

\medskip

After trigger registration for each function, the function can be invoked and the orchestrator function could decide to suspend itself. It will be later activated when the trigger fires.

To ensure that this PyWren code can be restarted and continue from the last point, we use \emph{event sourcing}. When the orchestrator code is launched,  an event sourcing action will re-run the code acquiring the results of functions from termination events. It will then be able to continue from the last point.

In our system prototype, the event sourcing is implemented in two different ways: native and external scheduler.

In the \emph{native scheduler}, the orchestration code is executed inside a Triggerflow Action. Our Triggerflow system enables then to upload  the entire orchestration code as an action that interacts with triggers in the system. When Triggerflow detects events that match a trigger, it awakens the native action. This code then relies on event sourcing to catch up with the correct state before continuing the execution.  In the native scheduler, the events can be retrieved efficiently from the context and thus accelerate the replay process. If no events are received in a period, the action will be scaled to zero. This guarantees reactive execution of event sourced code.

\begin{figure}[t!]
	\includegraphics[width=0.35\textwidth]{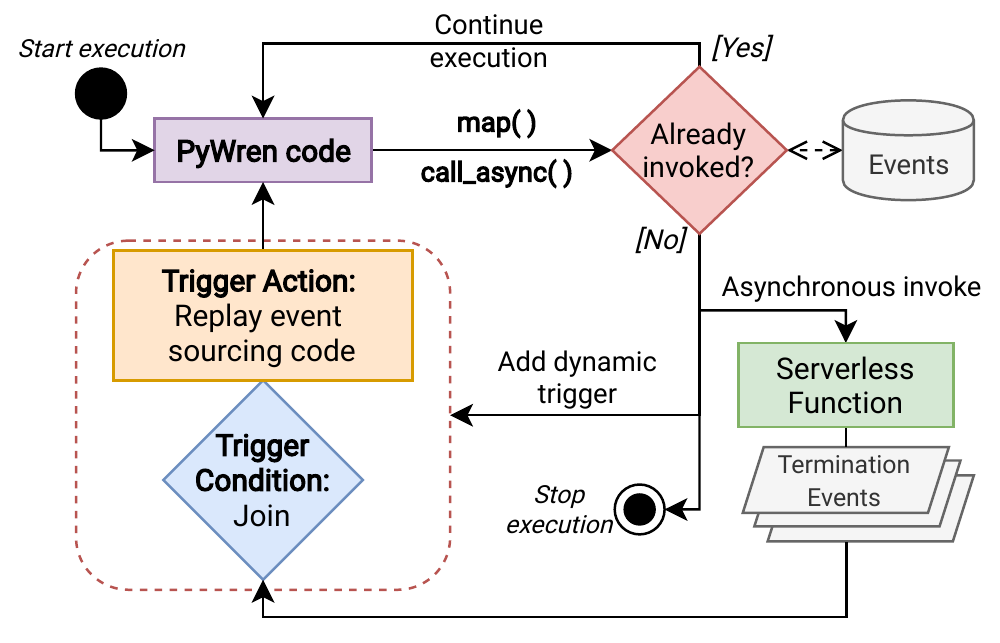}
	\centering
	\caption{Life cycle of an event sourcing-enabled workflow as code with IBM-PyWren as external scheduler.}
	\label{fig:event-source-pywren}
	\vspace{-8pt}
\end{figure}

In the \emph{external scheduler}, we use IBM PyWren \cite{ibmpy}, where the orchestration code is run in an external system, like a Cloud Function. Then, thanks to our Triggerflow service, the function can stop its execution each time it invokes for example a \emph{map()}, recovering their state (event sourcing) when it is awaken by our TF-Worker once all \emph{map()} function activations finished their execution. Moreover, to use our event sourcing version of PyWren, it is not required any change in the user's code. This means that the code is completely portable between the local-machine and the Cloud, so users can decide where to run their PyWren workflows without requiring any modification. The life cycle of a workflow using an external scheduler can be seen in Figure \ref{fig:event-source-pywren}.

\section{Validation}
\label{sec:validation}

Our experimental testbed consists of $5$ client machines with $4$ CPUs and $16$ GB RAM. On the server side, we deploy Triggeflow on a Kubernetes installation (v1.17.3) in a rack of $5$ Dell PowerEdge R$430$ (2 CPUs Intel(R) Xeon(R) CPU E5-2620 v4 @ 2.10GHz - 8 Cores/CPU - 32 Logical processors) machines with $16$GB RAM. All of these machines, including the clients, are connected via 10GbE network, and run Ubuntu Server $19.04$. For the experiments we use Kafka 2.4.0 (Scala 2.13), Redis 5.0.7, KEDA 1.3.0 and Knative 0.12.0.

\subsection{Load test}

The load test objective is to demonstrate that our system can support high-volume event processing workloads in an efficient way. This is mandatory if we want to support the execution of high performance scientific workflows. 

For the first experiment, we want to measure how many events per second can be processed by a worker that filters events from a message broker like Kafka or Redis Streams. Tables \ref{load_test_redis} and \ref{load_test_kafka} show the time to process $200$K events in a container using different CPU resources ($0.25$, $0.5$, $1$ and $2$). \emph{Noop} means that the worker is not doing any operation on the event. \emph{Join} refers to aggregated filters that process joining for different map jobs with $2000$ functions each. As we can see, the performance numbers tell that the system can process thousands of events per second. 

\begin{figure}[t!]
    \vspace{-12pt}
	\includegraphics[width=0.45\textwidth]{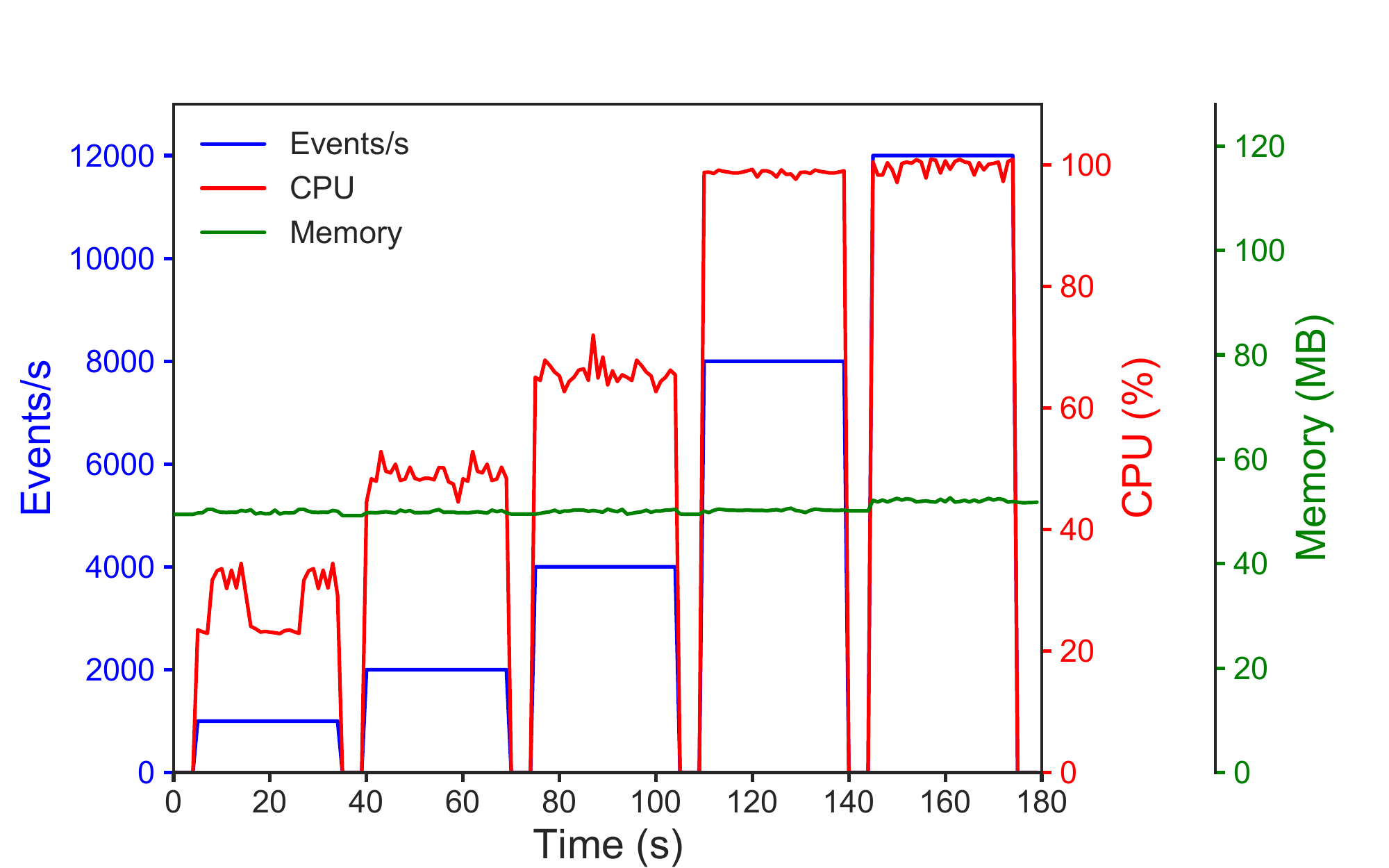}
	\centering
	\caption{Resource utilization depending on incoming number of events/second (1 Core w/ Redis)}
	\label{fig:load_test}
	\vspace{-8pt}
\end{figure}

The second experiment consists of measuring the actual resource usage (CPU and mem) of $1$ Core worker using Redis by injecting different numbers of events per second (form $1$K e/s to $12$K e/s). Figure \ref{fig:load_test} shows that, with a constant memory footprint, the CPU resource can cope with increasing number of events per second.

\begin{table}[h!]
\renewcommand{\arraystretch}{1.4}
\centering
\newcolumntype{P}[1]{>{\raggedleft\arraybackslash}p{#1}}
\caption{Maximum number of processed events/second using Redis Streams}
\begin{tabular}{|c|c|c|c|c|c|}
\hline
\textbf{Cores} & \textbf{Ev.} & \textbf{Noop (s)} & \textbf{Noop (e/s)}  & \textbf{Join (s)} & \textbf{Join (e/s)} \\ \hline\hline
\texttt{0.25}  & \texttt{200K} & \texttt{56.09}  &  \texttt{3565} & \texttt{59.83}  &  \texttt{3342}  \\ \hline
\texttt{0.5}   & \texttt{200K} & \texttt{28.03}  &  \texttt{7135} & \texttt{30.25}  &  \texttt{6611}  \\ \hline
\texttt{1}     & \texttt{200K} & \texttt{14.17}  &  \texttt{14114} & \texttt{14.56}  &  \texttt{13736} \\ \hline
\texttt{2}     & \texttt{200K} & \texttt{11.48}  &  \texttt{17421} & \texttt{12.02}  &  \texttt{16638} \\ \hline
\end{tabular}
\label{load_test_redis}
\end{table}

\vspace{-12pt}

\begin{table}[h!]
\renewcommand{\arraystretch}{1.4}
\centering
\newcolumntype{P}[1]{>{\raggedleft\arraybackslash}p{#1}}
\caption{Maximum number of processed events/second using Kafka}
\begin{tabular}{|c|c|c|c|c|c|}
\hline
\textbf{Cores} &\textbf{Ev.}& \textbf{Noop (s)} & \textbf{Noop (e/s)}  & \textbf{Join (s)} & \textbf{Join (e/s)}       \\ \hline\hline
\texttt{0.25}  & \texttt{200K} & \texttt{43.89}  &  \texttt{4556} & \texttt{49.30}  &  \texttt{4056}  \\ \hline
\texttt{0.5}   & \texttt{200K} & \texttt{18.01}  &  \texttt{11104} & \texttt{23.99}  &  \texttt{8336}  \\ \hline
\texttt{1}     & \texttt{200K} & \texttt{9.34}  &  \texttt{21413} & \texttt{11.31}  &  \texttt{17683} \\ \hline
\texttt{2}     & \texttt{200K} & \texttt{5.68}  &  \texttt{35211} & \texttt{7.56}  &  \texttt{26450} \\ \hline
\end{tabular}
\label{load_test_kafka}
\end{table}

\subsection{Auto-scaling}
In this case, the objective is to demonstrate that  TF-Workers can scale up and down based on the current active workflows.
We demonstrate here that our Triggerflow implementation on top of Kubernetes and KEDA can auto-scale on demand based on the number of events received in different workflows.

For this experiment, we use the entire testbed described above, and set the TF-Worker to use $0.5$ CPUs and $256$ MB of RAM. The test consists of $100$ synthetic workflows that send events during some arbitrary seconds, pause the workflow for a while (simulating a long-running action), then resume sending events, and finally stop the workflow. The test works as follows: It first starts $50$ workflows at a constant rate of $2$ workflows per second), after $100$ seconds it starts another $50$ workflows at a rate of $3$ workflows per second, and finally, after $70$ seconds, it starts $15$ more workflows at a rate of also $3$ workflows per second.

\begin{figure}[t!]
	\vspace{-12pt}
	\includegraphics[width=0.45\textwidth]{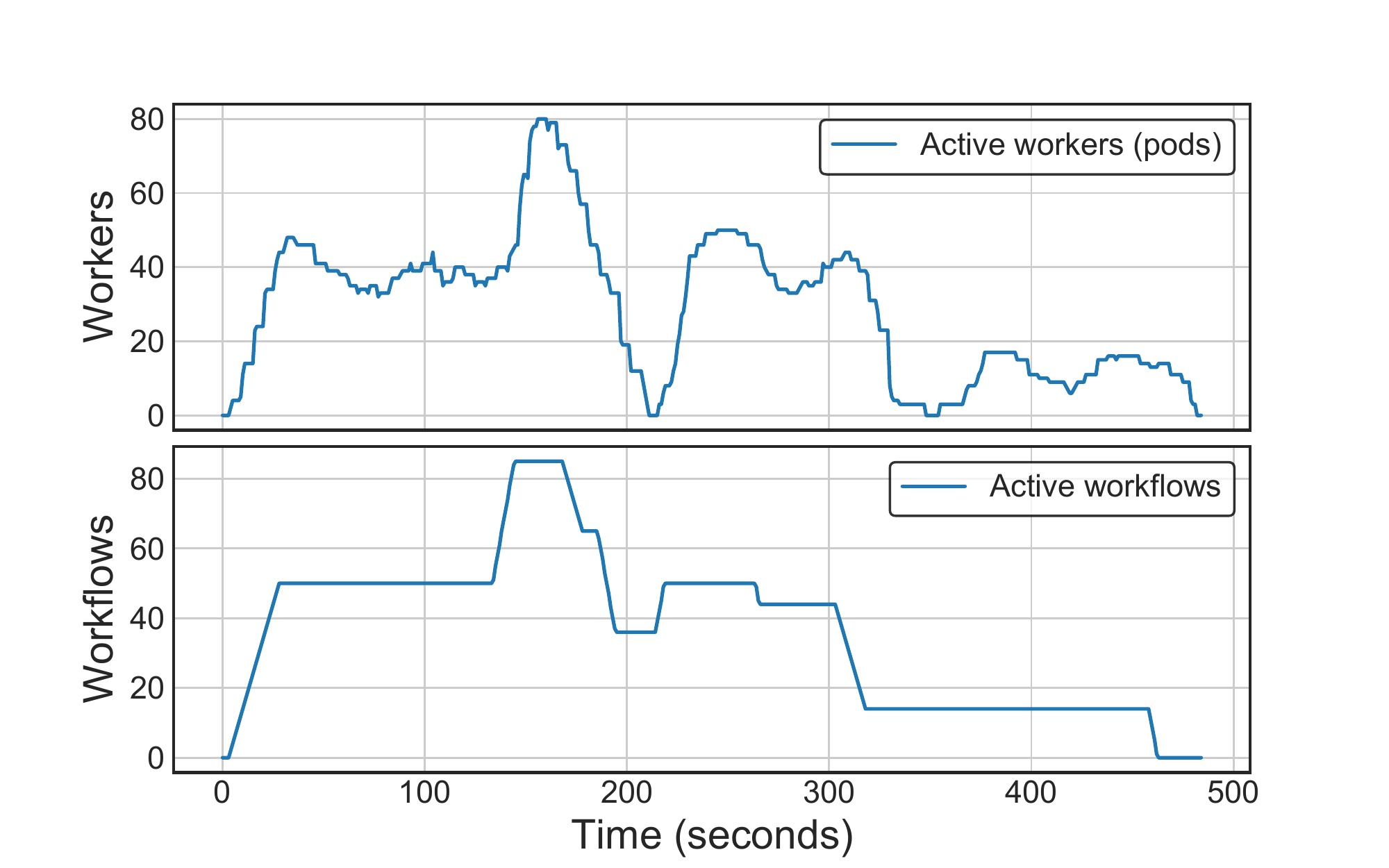}
	\centering
	\caption{TF-Worker auto-scaling test using KEDA}
	\label{fig:autoscaling}
	\vspace{-8pt}
\end{figure}

The results are depicted in Figure \ref{fig:autoscaling}. It shows how the TF-Workers scale up when the workflows start to send events, and scale down, even to zero (second 210 and 250), when the active workflows do not produce any event due to a long-running action. We can see how Triggerflow leverages the KEDA auto-scaler to activate or passivate workflows. Triggerflow is automatically providing fault tolerance, event persistence, and context and state recovery each time a workflow is resumed.

\subsection{Completion time and overhead}

The validation in this section demonstrates that Triggerflow shows comparable overhead to public Cloud orchestration systems.
We must be fair here: we are comparing an implementation of Triggerflow over dedicated and idle resources in our rack against public multi-tenant cloud services that may be used by thousands of users. The objective is not to claim that our system is better than them, but only to demonstrate that we can reach comparable overhead and performance.
Furthermore, most cloud orchestration systems are not designed for highly concurrent and parallel jobs, which can limit their performance in those scenarios.

We evaluate the run-time overhead of Amazon's, IBM's, and Microsoft's orchestration services. We consider as \emph{overhead} all the time spent outside the functions being composed, which~is easy to measure
in all platforms. For a sequential composition $g$ of $n$ functions  $g = f_1  \circ f_2 \circ \dots \circ  f_n$, it is just: 
\[
\small
\mbox{overhead }({g})=  \mbox{exec\_time}( g) -   \sum^n _{i = 1} \mbox{exec\_time}( f_i).
\]

It is important to note that our overhead definition includes the delays between function invocations, and the execution time of the orchestration function (for IBM Composer and ADF) or the delays between state transitions (for ASF). In the case of Triggerflow, the overhead depends on all the services in the architecture---i.e., latency to access Kafka or Redis, latency to invoke functions in IBM CF, etc.

For all the tests, we use a single TF-Worker with $0.5$ CPU Cores and $64$MB of RAM, and we list only the results when functions are in warm state. This implies that we do not consider the cold start of spawning the function containers and VMs. Our focus is on measuring the overhead of running function compositions. All the tests are repeated $10$ times. The results displayed are the median of those 10 samples and the standard deviation for the error intervals. Measurements are done during March of 2020. For IBM Cloud Functions (IBM CF) and AWS Lambda executions, we use the Python 3.8 runtime. The exception is Azure, which does not currently support Python for ADF, but C\#. The orchestration functions are implemented in the default language available in each platform: Node.js for IBM Composer, and C\# for ADF. ASF orchestration is specified in Amazon States Language (JSON-based format) using the console editor.

For the \emph{sequential workflows}, we quantify the overhead for sequential compositions of length \emph{n} in \{5, 10, 20, 40, 80\}. For simplicity, all the functions in the sequence are the same: a function that sleeps for 3s, and then returns. For the \emph{parallel workflows}, we define a workflow with a single parallel stage composed of $n$ parallel instances of the same task, with $n$ ranging from $5$ to $320$, and doubling each time. This task has a fixed duration of $20$ seconds. Consequently, any execution of the experiment should ideally last $20$ seconds, irrespective of $n$ or the environment. To put it in another way, in an ideal system with no overhead, the execution time of the $n$ concurrent tasks should match that of a single task. Therefore, we compute the overhead of the orchestration system by subtracting the fixed time of a single task, namely $20$ seconds, from the total execution time.

\subsubsection{DAGs and State Machines}
For the DAG and State Machine use cases, we evaluated our DAG engine interface against IBM Composer, AWS Step Functions, AWS Step Functions Express, and Azure Durable Functions. It is important to state that these results are exactly the same we would get for the State Machine implementation over Triggerflow. Sequences and parallel jobs in state machines and DAGs use the same triggers.

\begin{figure}[t!]
	\includegraphics[width=0.45\textwidth]{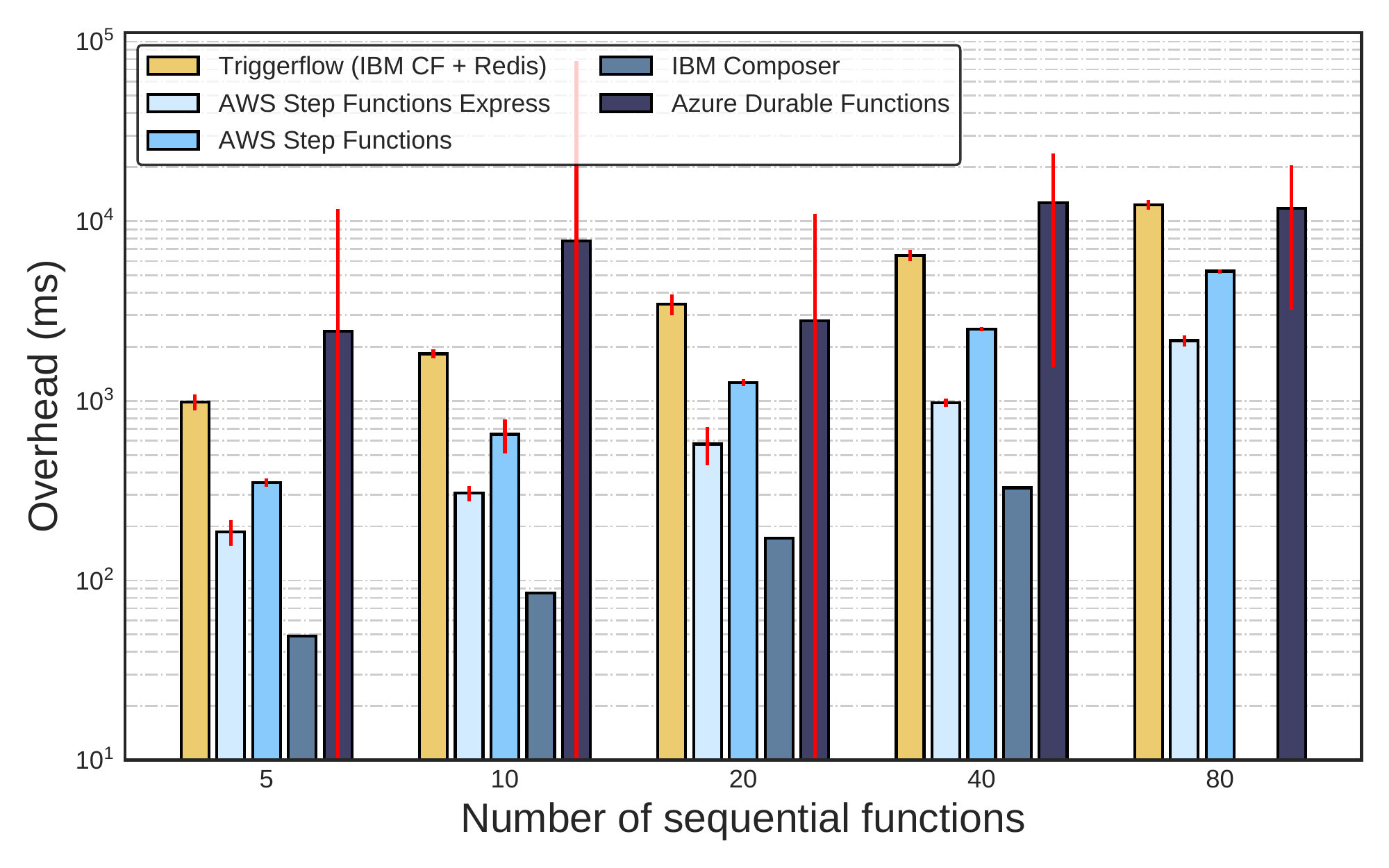}
	\centering
	\caption{DAG overhead comparison for sequences}
	\label{fig:dag_sequence_plot}
	\vspace{-8pt}
\end{figure}

\medskip
\noindent{\textbf{Sequential workflows.}} The resultant overhead is represented in Figure \ref{fig:dag_sequence_plot}. In general, Triggerflow's overhead is higher than in other orchestration systems. In this case, almost all overhead comes from the IBM Cloud Functions invocation latency using its public API, which is about $0.13$s. When multiplied by 80 functions, it adds up to approximately 10 seconds of overhead. Amazon Step Functions, however, can use internal trigger protocols rather than the public API, which explains the lower invocation latency. In addition, it seems that using Express Workflows does not provide a considerable speed improvement compared to regular ASF when using sequential workloads, so they are probably not worth the extra cost for this kind of job. IBM Composer is the fastest in sequences, but with the drawback of its limitation of only 50 transitions per composition. Finally, Azure Durable Functions present competent overheads, although being quite unstable for short sequences. This is probably because ADF is designed and optimized for long-running sequential workloads.

\begin{figure}[t!]
	\includegraphics[width=0.45\textwidth]{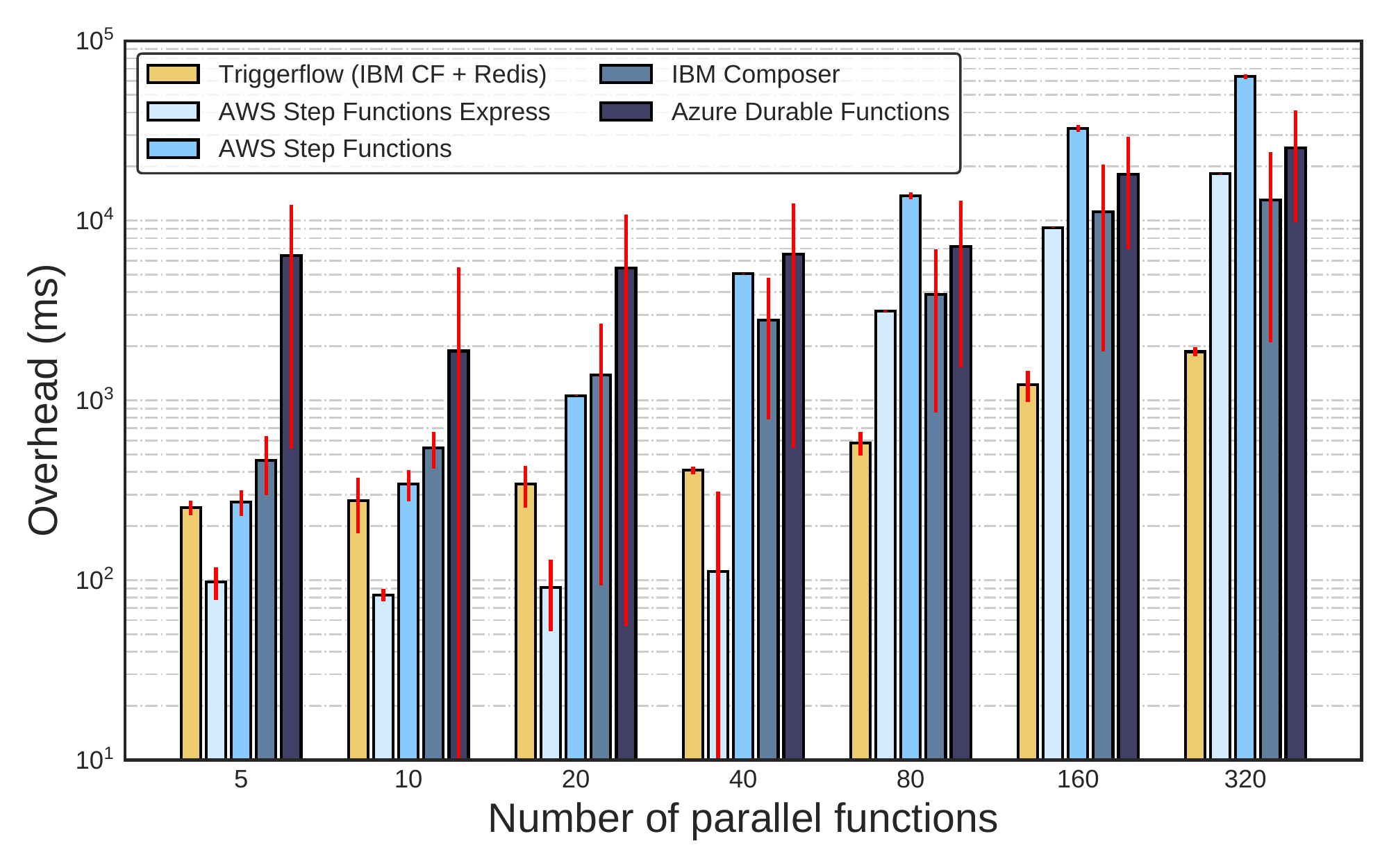}
	\centering
	\caption{DAG overhead comparison for parallel workflows}
	\label{fig:dag_map_plot}
	\vspace{-8pt}
\end{figure}

\medskip
\noindent{\textbf{Parallel workflows.}} For small-sized compositions (5 to 10), we can see in Figure \ref{fig:dag_map_plot} that Triggerflow and AWS Step Functions yield similar overhead, both being outperformed by Express Workflows nonetheless. Express Workflows has a wider range of error though, while regular Step Functions, Triggerflow and IBM Composer are more stable. Express Workflows perform similarly regardless of the number of parallel functions until it reaches about 80, when its performance drops drastically and the overhead skyrockets for no apparent reason. From 80 functions and up, Express Workflows and IBM Composer have similar overheads.

From 80 parallel functions and up, we also see that Triggerflow has the lowest overhead, proving that event-driven function composition is indeed well suited for parallel function joining.

Azure Durable Functions yield the worst results when used for small-sized function joining and is considerably unstable. However, it turns to be equivalent to the other orchestration systems when joining a higher number of concurrent functions.

\subsubsection{Workflow as Code and Event Sourcing}
The objective here is to evaluate Workflow as Code and event sourcing overheads in Triggerflow compared to Azure Durable Functions. We compare both sequential and parallel constructs.

For the event sourcing use case, we evaluate both the \emph{external scheduler} (IBM-PyWren) and the \emph{native scheduler} (Triggerflow action).
One the one hand, we measure and compare the performance of our modified version of IBM-PyWren for Triggerflow with the original version of IBM-PyWren (\emph{external scheduler}). In this case we evaluate 4 different scenarios: 1) The original IBM-PyWren, which makes use of IBM Cloud Object Storage (COS) to store the events and results. 2) The modified version of IBM-PyWren for Triggerflow that stores the results in COS (original IBM-PyWren behavior), but sends the termination events trough a Redis Stream. 3) The Triggerflow IBM-PyWren that sends the events and results trough a Kafka Topic. And 4) the Triggerflow IBM-PyWren that sends the events and results trough a Redis Stream.

On the other hand, we evaluate the native Triggerflow event sourcing scheduler, where the orchestration code is executed as part of the trigger action. In this case we compare the results against the Azure Durable Functions (ADF) service, which is the only FaaS workflow orchestration service that employs an event sourcing technique to execute the workflows.

\begin{figure}[t!]
	\includegraphics[width=0.45\textwidth]{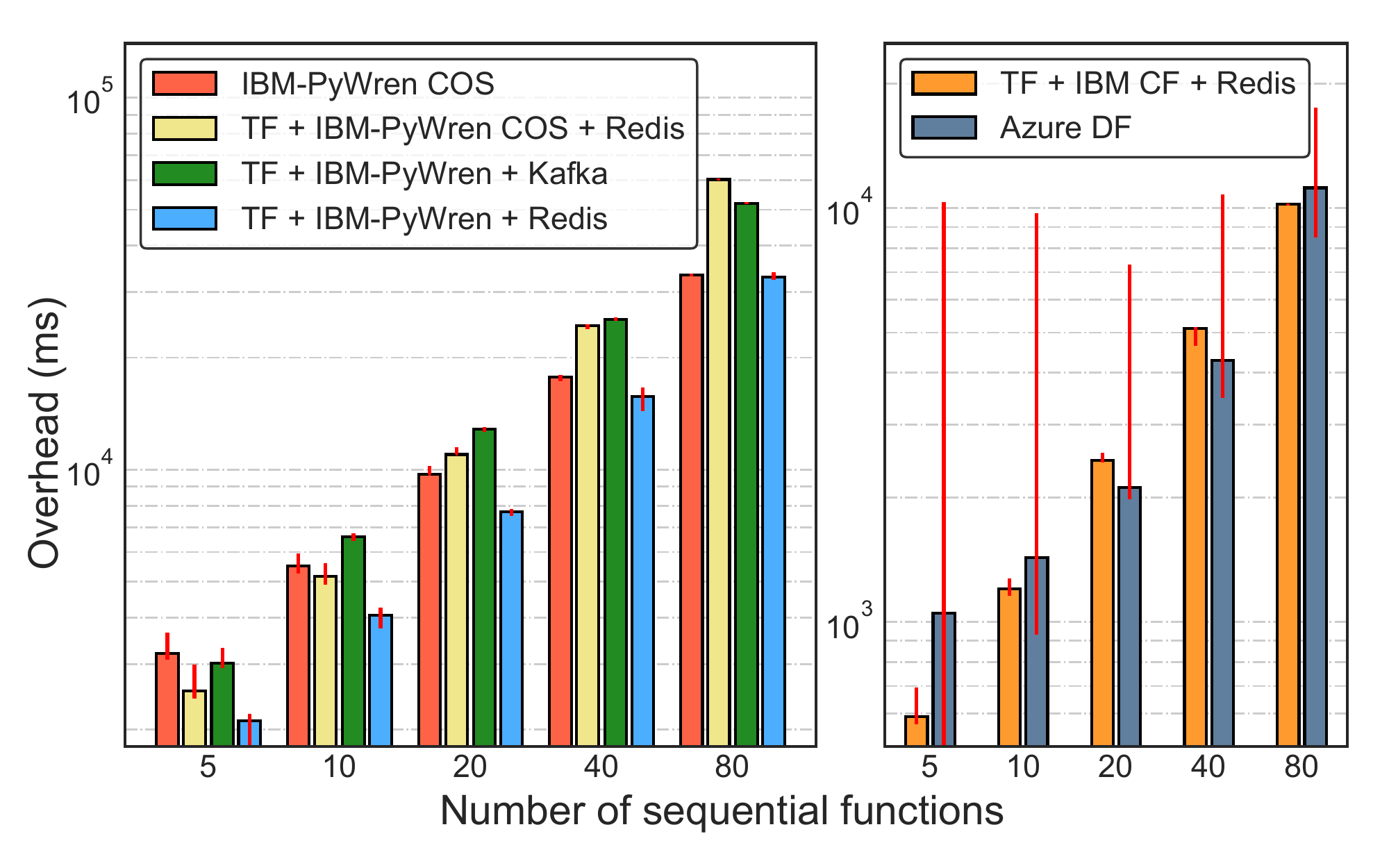}
	\centering
	\caption{Event sourcing overhead comparison for sequences. PyWren vs TF-PyWren on the left side. Triggerflow vs Azure Durable Functions on the right side.}
	\label{fig:pw_adf_sequence}
	\vspace{-8pt}
\end{figure}

\medskip
\noindent{\textbf{Sequential workflows.}} Figure \ref{fig:pw_adf_sequence} shows the overhead evolution when increasing the length of the sequence. The overhead added by both the native and external schedulers grows up linearly based on the number of functions in the sequence. As we can see, the results are very stable, meaning that the behavior is implementation-related, and not a problem with resources.

For the \emph{external scheduler}, we can see comparable performance between the original IBM-PyWren and our modified version for Triggerflow. Overhead evolves similarly in all scenarios. PyWren has to serialize and upload the function and the data to COS before executing it, creating overhead common for all scenarios. The remaining overhead comes from the place and the way these events are retrieved to recover the state of the execution (event sourcing). This means that the event source service---either COS, Kafka, or Redis---, has direct impact on these results.  For example, the main drawback of using COS in both the original (1) and Triggerflow (2) versions of IBM-PyWren is that they have to individually download the results from COS. This fact substantially increases the total time needed to execute a workflow, since for each step it has to retrieve all the previous events. In this case, for a workflow with $n$ steps, IBM-PyWren has to perform a total of $n(n+1)/2$ requests. In contrast, in the scenarios where IBM-PyWren does not use COS, and stores the events in a Kafka Topic (3) or a Redis Stream (4), it only needs one request to retrieve all the events in each step. Then, it only needs $n$ requests to these services to complete the execution of a workflow. If we compare scenarios 2 and 3, we see better performance if we use a Redis Stream instead of a Kafka Topic. This is mainly caused by the Kafka library, which adds a fixed overhead of $0.25$s each time the orchestration function is awaken and creates a consumer. This means that using a Kafka Topic as event store has a fixed overhead of $n*0.25$ seconds.

For the Triggerflow \emph{native scheduler}, it is important to note that the functions are already deployed in the cloud (in contrast with PyWren that has to serialize and upload them each time). Moreover, the orchestration code is execute within the TF-Worker that contains all the events loaded in memory, so it does not need to retrieve them from the event source (Kafka, Redis) in each step. Compared to ADF, we obtain similar overhead. As stated in the previous section, the overhead comes mainly from the fact that invoking an action in IBM CF service takes around $0.13$s. This means that, for a workflow of $n$ steps, Triggerflow has a fixed overhead of $n*0.13$ when using IBM CF.

\begin{figure}[t!]
	\includegraphics[width=0.45\textwidth]{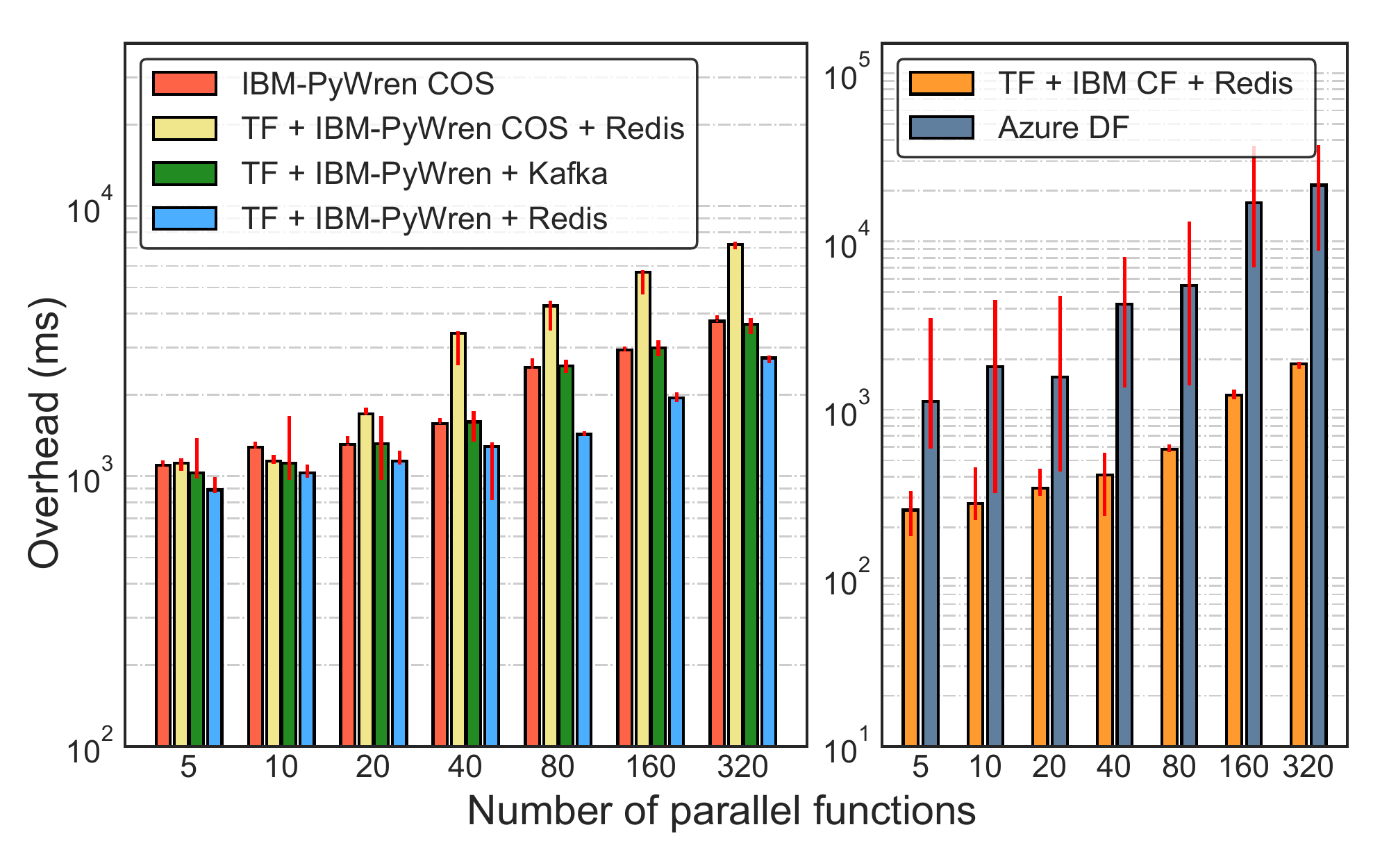}
	\centering
	\caption{Event sourcing overhead comparison for parallel workflows. PyWren vs TF-PyWren on the left side. Triggerflow vs Azure Durable Functions on the right side.}
	\label{fig:pw_adf_parallel}
	\vspace{-8pt}
\end{figure}

\medskip
\noindent{\textbf{Parallel workflows.}} For this experiment, we evaluate the same scenarios described above. The results are depicted in Figure \ref{fig:pw_adf_parallel}. In this case, for the \emph{external scheduler}, the original IBM-PyWren and the Triggerflow IBM-PyWren version have also similar overhead, being scenario 4---which uses Redis as event store---the best approach. In the Kafka scenario (3), the overhead of $0.25$s described above is negligible, since in this experiment the orchestration function is awaken only once. The main difference in the performance between scenarios 1 and 2 is that the original IBM-PyWren is running all the time and polling the results as they are produced. In contrast, in the Triggerflow version of IBM-PyWren that uses COS (2), the TF-Worker first waits for all activations to finish to awake the orchestration function, that then has to retrieve all the events and results from COS. Finally, with the \emph{native scheduler}, Triggerflow is faster for parallel workflows compared to ADF. 

\subsection{Scientific Workflows}

\begin{figure}[t!]
	\includegraphics[width=0.45\textwidth]{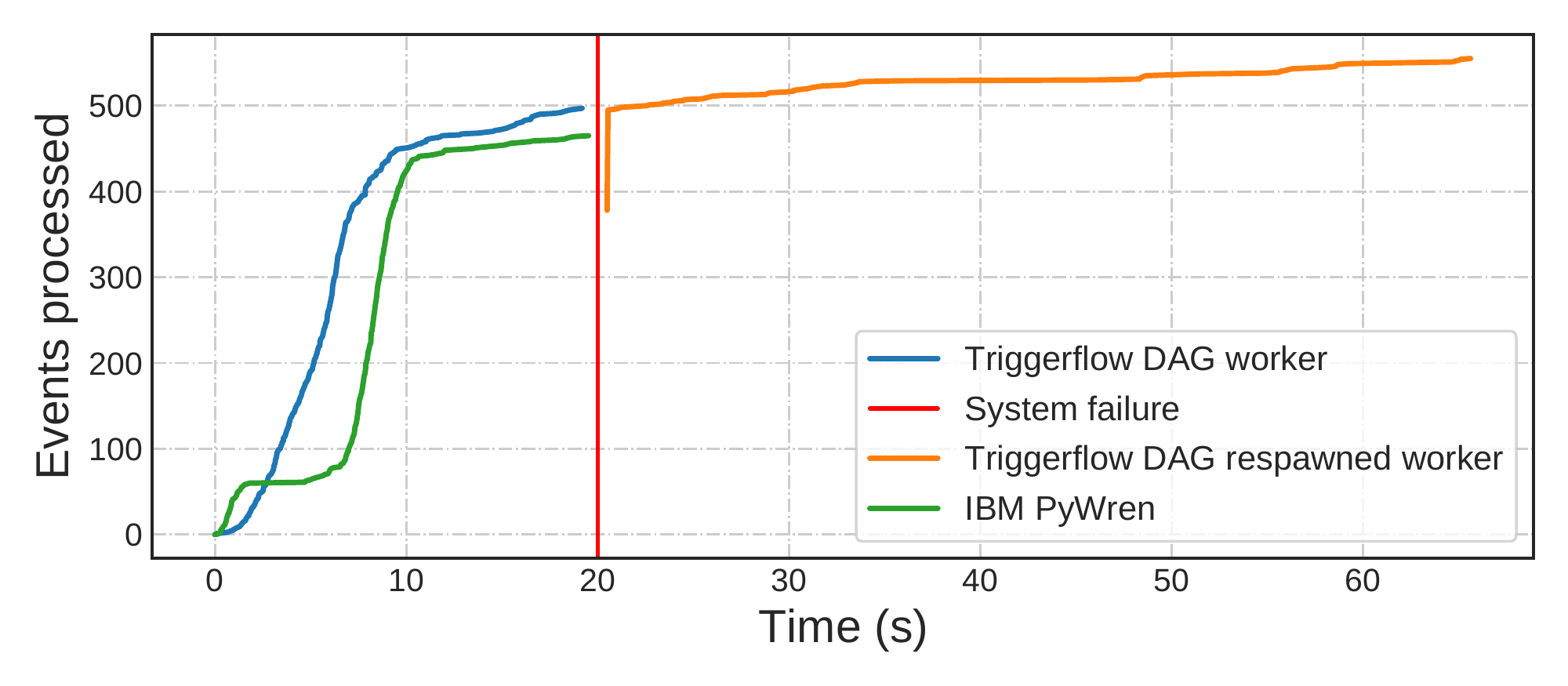}
	\centering
	\caption{Scientific workflow execution progression over time, with an intended system failure at the 20th second.}
	\label{fig:fault_tolerance}
	\vspace{-8pt}
\end{figure}

We adapted a geospatial scientific workflow, that was originally implemented with IBM-PyWren, to work with our DAGs interface. The objective of the workflow is to compute the evapotranspiration and thus the water consumption of the crops from a set of a partitioned geographical region using the Penman-Monteith equation. Due to the nature of the workflow, and despite the optimizations applied, the workflow's execution time is similar to that provided by IBM-PyWren. The main difference lies in the workflow programming model: DAGs are more geared towards dissecting the workflow into independent tasks and their dependencies, while PyWren opts for a map-reduce model. An important point in favor of Triggerflow is its automatic and transparent fault tolerance provided by the event source and trigger persistent storage. Figure \ref{fig:fault_tolerance} depicts the progression of a workflow run of the scientific workflow, using Kafka as the event source and Redis for the trigger storage. To check the system's fault tolerance, we intentionally stopped the execution of the Triggerflow worker and the IBM-PyWren execution in the 20th second of the workflow execution. Triggerflow rapidly recovers the trigger context from the database and the uncommitted events from the event source, and finishes its execution correctly. In contrast, IBM-PyWren stops and loses the state of the workflow, having to re-execute the entire workflow wasting time and resources.

\begin{figure}[t!]
	\includegraphics[width=0.45\textwidth]{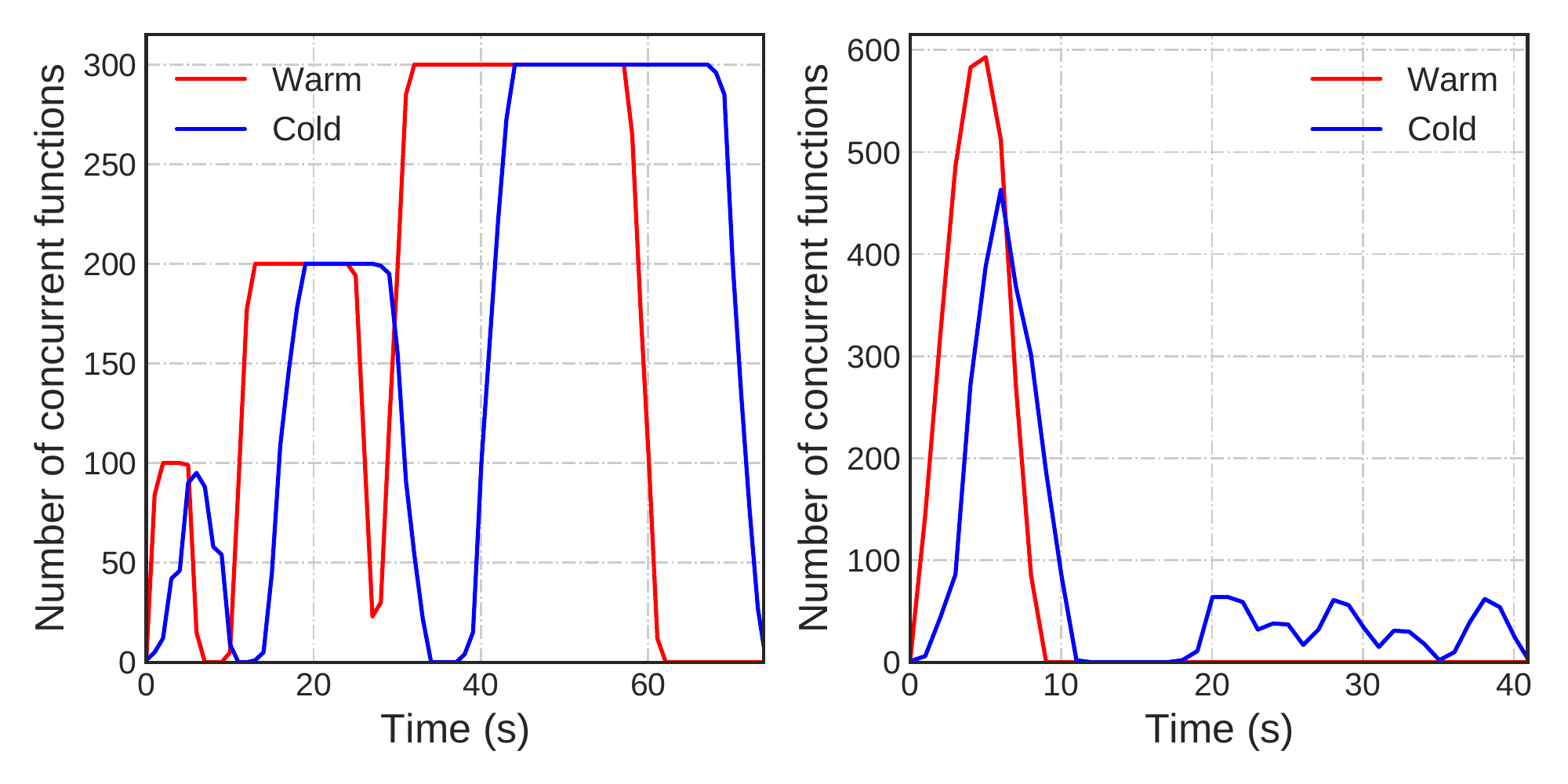}
	\centering
	\caption{(a) Parallelism and total execution time in a sequential workflow that increases its map size and execution time every step. (b) Parallelism and total execution time of a single map task with high concurrency.}
	\label{fig:prewarm}
	\vspace{-8pt}
\end{figure}

To demonstrate Triggerflow's ability to introspect triggers with its Rich Trigger API, we have also implemented a service over the DAGs interface that automatically and transparently prewarms function containers on IBM Functions to increase the efficiency and overall parallelism, reduce total execution time and mitigate straggler functions effects in workflows that require high performance and throughput. Figure \ref{fig:prewarm} shows its effects. Thanks to Triggerflow's interception mechanisms we can also transparently apply other data pre-fetching optimizations in scientific workflows.

\section{Conclusions}

We have presented in this paper Triggerflow: a novel building block for controlling the life cycle of Cloud applications. As more applications are compiled to the Cloud, our system permits to encode their execution flow as reactive triggers in an extensible way. The novelty of our approach relies on four key aspects: serverless design, extensibility, support for heterogeneous workflows, and performance for high-volume workloads.

TriggerFlow can become an extensible control plane for deploying reactive applications in the Cloud.  We implemented and validated different orchestration systems based on State Machines (ASF), Directed Acyclic Graphs (Airflow), and Workflow as Code (PyWren).


Finally, the major limitations about TriggerFlow as an event-based orchestration system are debuggability and developer experience. As an open source project,  TriggerFlow would clearly benefit from tools and user interfaces to simplify the overall observability and life-cycle support of the system. 

\begin{acks}
This work has been partially supported by the EU Horizon 2020 programme under grant agreement No 825184.
\end{acks}

\bibliographystyle{ACM-Reference-Format}
\bibliography{debs}



\end{document}